%% file: bare_jrnl.tex
\documentclass[journal]{IEEEtran}
\usepackage[pagebackref=true,breaklinks=true,letterpaper=true,colorlinks,bookmarks=false]{hyperref}

\usepackage{cite}

\ifCLASSINFOpdf
    \usepackage[pdftex]{graphicx}
    \graphicspath{{../pdf/}{../jpeg/}}
    \DeclareGraphicsExtensions{.pdf,.jpeg,.png}
\else
\fi
\usepackage{amsmath}
\usepackage{amssymb}
\usepackage{url}
\usepackage{booktabs}
\usepackage{graphicx}
\usepackage{algorithm}
\usepackage{color}
\usepackage{algorithmic}
\usepackage[pagebackref=true,breaklinks=true,letterpaper=true,colorlinks,bookmarks=false]{hyperref}

\begin{document}
%
\title{Coupled-Projection Residual Network for MRI Super-Resolution}
%
%
%
\author{Chun-Mei Feng,
	Kai Wang,
	Shijian Lu,
	Yong Xu,~\IEEEmembership{Senior~Member,~IEEE},
	Heng Kong
	and Ling Shao,~\IEEEmembership{Senior~Member,~IEEE}
\thanks{C.-M. Feng is with the Bio-Computing Research Center, Harbin Institute of Technology, Shenzhen 518055, China (e-mail: fengchunmei0304@foxmail.com)}

\thanks{K. Wang is with Shenzhen Institutes of Advanced Technology, Chinese Academy of Sciences, Shenzhen, China(e-mail: kai.wang@siat.ac.cn).}
\thanks{S. Lu is with the School of Computer Science and Engineering, Nanyang
Technological University, Singapore 639798 (e-mail: shijian.lu@ntu.edu.sg).}
\thanks{Y. Xu (Corresponding Author) is with the Bio-Computing Research Center,  Harbin Institute of Technology, Shenzhen 518055, China, and also with the Key Laboratory of Network Oriented Intelligent Computation, Shenzhen 518055, China (e-mail: yongxu@ymail.com).}
\thanks{H. Kong is with Shenzhen Univerisity General Hospital, Shenzhen University, Shenzhen 518055, China(e-mail: generaldoc@126.com).}
\thanks{L. Shao is with the Inception Institute of Artificial Intelligence, Abu Dhabi, United Arab Emirates (e-mail: ling.shao@ieee.org).}
}

%
%

\markboth{Journal of \LaTeX\ Class Files,~Vol.~14, No.~8, August~2015}%
{Shell \MakeLowercase{\textit{et al.}}: Bare Demo of IEEEtran.cls for IEEE Journals}
%



\maketitle

\begin{abstract}
Magnetic Resonance Imaging (MRI) has been widely used in clinical application and pathology research by helping doctors make more accurate diagnoses. On the other hand, accurate diagnosis by MRI remains a great challenge as images obtained via present MRI techniques usually have low resolutions. Improving MRI image quality and resolution thus becomes a critically important task. 
This paper presents an innovative Coupled-Projection Residual Network (CPRN) for MRI super-resolution. The CPRN consists of two complementary sub-networks: a shallow network and a deep network that keep the content consistency while learning high frequency differences between low-resolution and high-resolution images. The shallow sub-network employs coupled-projection for better retaining the MRI image details, where a novel feedback mechanism is introduced to guide the reconstruction of high-resolution images. The deep sub-network learns from the residuals of the high-frequency image information, where multiple residual blocks are cascaded to magnify the MRI images at the last network layer. Finally, the features from the shallow and deep sub-networks are fused for the reconstruction of high-resolution MRI images. For effective fusion of features from the deep and shallow sub-networks, a step-wise connection (CPRN\_S) is designed as inspired by the human cognitive processes (from simple to complex). Experiments over three public MRI datasets show that our proposed CPRN achieves superior MRI super-resolution performance as compared with the state-of-the-art. Our source code will be publicly available at \href{http://www.yongxu.org/lunwen.html}{http://www.yongxu.org/lunwen.html}.
\end{abstract}

%
\IEEEpeerreviewmaketitle

\section{Introduction}
\label{intro}
\input{introduction}

\section{Related work}
\label{related}
\input{relatedwork}
\section{Methodology}
\label{method}
\input{ourwork}

\section{Experiments}
\label{experiment}
\input{experiments}

\section{Conclusion}
\input{conclusion}

\section{Acknowledge}
This work was supported in part by the Natural Science Foundation of China under Grant 61573248, Grant 61802267 and Grant 61732011, in part by the Natural Sci-ence Foundation of Guangdong Province (Grant 2017A030313367), and in part by the Shenzhen Municipal Science and Technology Innovation Council under Grant JCYJ20180305124834854 and JCYJ20160429182058044.

\IEEEtriggercmd{\enlargethispage{-5in}}


%
\bibliographystyle{IEEEtran}
\bibliography{IEEEabrv,yinyong}
\end{document}

%% file: introduction.tex
\IEEEPARstart{I}mage super-resolution is currently a hotspot issue in natural imaging \cite{3,2,1}, medical imaging \cite{tip_4}, surveillance \cite{7,6} and security \cite{8}. It allows to recover High-Resolution Images ($I^{HR}$) with better visual quality and refined details from the corresponding Low-Resolution Images $I^{LR}$. Magnetic Resonance Imaging (MRI) provides powerful support for disease diagnosis and treatment \cite{10,9}. On the other hand, high resolution MRI is expensive and subject to artifacts due to the elaborate hardware \cite{11}. Moreover, the longtime data acquisition and breath holding, combined with the unconscious or autonomous movement of patients, often lead to the missing of key information and motion artifacts in images \cite{12}. Super-resolution reconstruction of MRI images thus offers new promises for mitigating the costs of high resolution MRI technology. It  simplifies the MRI scanning process effectively, shortens the scanning time, reduces the use of MRI contrast agents, and is also safer for patients \cite{13}. By restoring high-resolution images, pathological lesions can be detected with high precision, enabling doctors to carry out more accurate diagnoses. Thus, the reconstruction of MRI data requires higher textural detail than traditional super-resolution tasks in clinical diagnosis \cite{15,14,5}.


Conventional image super-resolution approach restores the original $I^{HR}$ by fusing multiple $I^{LR}$ of the same scene. It is an ill-posed inverse problem \cite{18} due to the deficiency of $I^{LR}$, ill-conditioned registration and the absence of blurring operators. Specifically, numerous $I^{HR}$ produce the same $I^{LR}$ after resolution degradation, making it difficult to restore the image details accurately \cite{19}. For medical images, even small changes in textural details can affect a doctor’s diagnose \cite{5}. Prior knowledge is therefore often exploited to normalize the Super-Resolution Image ($I^{SR}$) generation process \cite{20}. In traditional methods, this prior information can be learned from several pairs of high and low-resolution images \cite{20}. In addition, different methods have been proposed to stabilize the inversion of the ill-posed problem, such as prediction-based method \cite{21}, edge-based method \cite{22} and sparse representation method \cite{23,20}. But these methods often over-smooth images because of ringing and jagged artifacts \cite{24,20}.

Deep learning can learn the mutual dependency between input and output data, which has been widely explore to restore the image details for precise super resolution \cite{27,26,25,16}. The deep learning-based super-resolution aims to directly learn the end-to-end mapping function of $I^{LR}$ to $I^{HR}$ through a neural network. Specifically, it extracts higher-level abstract features by multi-layer nonlinear transformation and learns the implying rules from data via the powerful capability of data fitting. Such learning capability empowers it to make reasonable decision or prediction for new data \cite{25}. The deep learning based super-resolution consists of 4 key steps: 1) Collect original images as $I^{HR}$ and obtain $I^{LR}$ by down-projection; 2) Feed the $I^{HR}$ into convolutional networks to extract feature; 3) Reconstruct the $I^{SR}$ by deconvolutional layer or up-projection; 4) Calculate the loss between $I^{SR}$ and $I^{HR}$ to optimize the super-resolution networks. 


Although great progress has been made in medical image super-resolution in recent years, several challenges remain. The first is about the curse of network layers - too few network layers often degrade the super-resolution performance due to the insufficient model capacity whereas too many layers often make optimization challenging and also introduce high computational costs \cite{25}. The second is about the monotonous structure of most existing network structures which makes it less effective to improve the super-resolution by increasing network layers. The third is about new noises which many existing super-resolution algorithms tend to introduce during up-projection processes. This directly leads to unrealistic image details that could mislead the doctor’s diagnosis seriously.

In this paper, we propose an innovative network for MRI super-solution. The contributions can be summarized in two major points. First, a Coupled-Projection Residual Network (CPRN), contains a shallow network and a deep network for effective MRI image super-solution. 
Specifically, the shallow network exploits coupled-projection to calculate the errors of repeated up-projection and down-projection for preserving more interrelations between $I^{LR}$ and $I^{HR}$. Such coupled-projection helps retain enriched image details as more as possible at the early stage of the network, and improves the alignment of the content of reconstructed images and the $I^{HR}$. The deep network inherits features of the shallow network but learns high-frequency differences between $I^{LR}$ and $I^{HR}$ by cascading the residuals. Second, a step-wise connection module termed by CPRN\_S is designed for refining the MRI image super-resolution. By fusing the feature maps from the down-projection of the shallow network and the corresponding output of the residual blocks, it improves the MRI image super-resolution effectively by relating the information from the deep and shallow networks and strengthening the feature propagation. Experiments show that CPRN\_S  saves up to 30\% of network parameters but achieves superior super-resolution performance, more detailes to be discussed in Experiments.

The rest of this paper is organized as follows. Section II briefly described related works on image super-resolution. Section III presents our proposed CPRN in details. Section IV then describes experimental results. Finally, a few concluding remarks are drawn in Section V.

%% file: relatedwork.tex

\subsection{DNN based Image Super-Resolution} 
Image super-resolution has been studied by interpolation and statistics based techniques \cite{32,34,33} in early years that aim to reconstruct $I^{HR}$ and restore image details and realistic textures. For example, the relationship between $I^{LR}$ and $I^{HR}$ can be learned from the correspondence function that is obtained via neighbor embedding and sparse coding techniques \cite{25,35,36,37,38}. In recent years, deep neural networks (DNNs) have been widely studied for image super-resolution and much improved super-resolution performance has been obtained \cite{40,39,30}. For example, \cite{25} proposes a Super-Resolution Convolutional Neural Network (SRCNN) that first uses bicubic interpolation to enlarge an $I^{LR}$ to the target size and then produces $I^{SR}$ via nonlinear mapping as fitted by a three-layer convolutional network. \cite{26} proposes the Faster-SRCNN that speeds up the SRCNN by adopting a smaller kernel size and sharing the mapping layers. \cite{27} proposes Efficient Sub-Pixel CNN (ESPCN) that reduce computational complexity by extracting features from $I^{LR}$ of original size directly. Based on CNN, Oktay et al. proposed de-CNN that improves reconstruction by using multiple images acquired from different viewing planes \cite{50}. SRCNN3D generates brain $I^{HR}$ from input $I^{LR}$ by using three-dimensional CNN (3DCNN) and the patches of other brain $I^{HR}$ \cite{28}. 
\cite{29} presents an Anti-Aliasing (AA) self-super resolution (SSR) algorithm that exploits high-frequency information of in-plane MRI slices and is capable of reconstructing $I^{SR}$ without external training data. 
On the other hand, all aforementioned methods are monotonous which miss to exploit features of various structures sufficiently. A multi-structure network is desired to capture richer features for more effective image super-resolution. 

Generative Adversarial Network (GAN) has recently been applied to different tasks such as image recognition, style transfer as well as super-resolution. \cite{30} first applies GAN for image super-resolution which labels the discriminator by $I^{HR}$ and feeds $I^{LR}$ to the generator to compute $I^{SR}$. \cite{31} presents a novel GAN for video super-resolution where the GAN is enhanced by a distance loss in feature and pixel spaces. \cite{16} combines GAN and 3DCNN for image super-resolution at multiple levels. \cite{51} presents a GAN-based cascade refinement network that integrates a content loss and an adversarial loss to reconstruct phase contrast microscopy images. Although GAN-based medical image super-resolution methods achieve very promising PSNR, they tend to hallucinate image fine details which is extremely unfavorable for medical images \cite{16}.

\subsection{Feedback Mechanism} 
The feedback mechanism decomposes the prediction process into multiple steps that implement feedback by estimating and correcting the current estimation  iteratively \cite{57,58,59,60}. It has been widely used in human pose estimation and image segmentation \cite{57,58}. For example, \cite{57} presents an Iterative Error Feedback (IEF) network that improving the initial solution gradually by back-progagating the error prediction and learning richer data representations. \cite{58} makes predictions for image segmentation by taking advantage of the implicit structure underlying the output space. \cite{48} applies feedback mechanism to super-resolution where a Deep Back-Projection Network (DBPN) was designed for projecting errors at each stage. Back-projection helps to reduce the reconstruction error effectively via up-sampling $I^{LR}$ and calculating reconstruction errors iteratively. 

Our proposed super-resolution network extends the feedback mechanism by introducing coupled-projection blocks which exploits alternative up-projection and down-projection for computing reconstruction errors and preserving more interrelations between $I^{LR}$ and $I^{HR}$. The content of reconstructed images can be aligned with the $I^{HR}$ by the coupled-projection.

\subsection{Skip-Connection} 
Skip-connection of network, such as residual-networks and dense-based networks, enables the flexible transmission, combination and reuse of features. For residual-based networks, Very  Deep  Super-Resolution (VDSR) exploits residual idea to include more network layers to expand the receptive field \cite{43}, where zero padding is implemented to keep the size of feature maps and final output images unchanged. \cite{43} and \cite{39} achieve multi-scale super-resolution within a single network and improve the computational efficiency greatly. \cite{39} presents an Enhanced Deep Super-Resolution network (EDSR) that removes redundant network modules and is more suitable for low-level computer vision problems. \cite{44} presents a Deeply-Recursive Convolutional Network (DRCN) that deepens the network structure by combining recursive neural networks and skip connection in residual learning. \cite{45} describes a symmetric encoder-decoder network (RED) that introduces a deconvolutional layer for each convolutional layer, where the convolutional layers capture abstract image contents and the deconvolutional layers enlarge feature sizes and restore image details. \cite{46} presents a Deep Recursive Residual Network (DRRN) that adopts multi-path recursive model by local and global residual learning and multi-weight recursive learning. \cite{47} presents a Laplacian Pyramid Super-Resolution Network (LapSRN) that can obtain intermediate reconstruction via progressive up-sampling to finer levels.

\begin{figure*}[h]
\centering
  \includegraphics[width=0.95\textwidth]{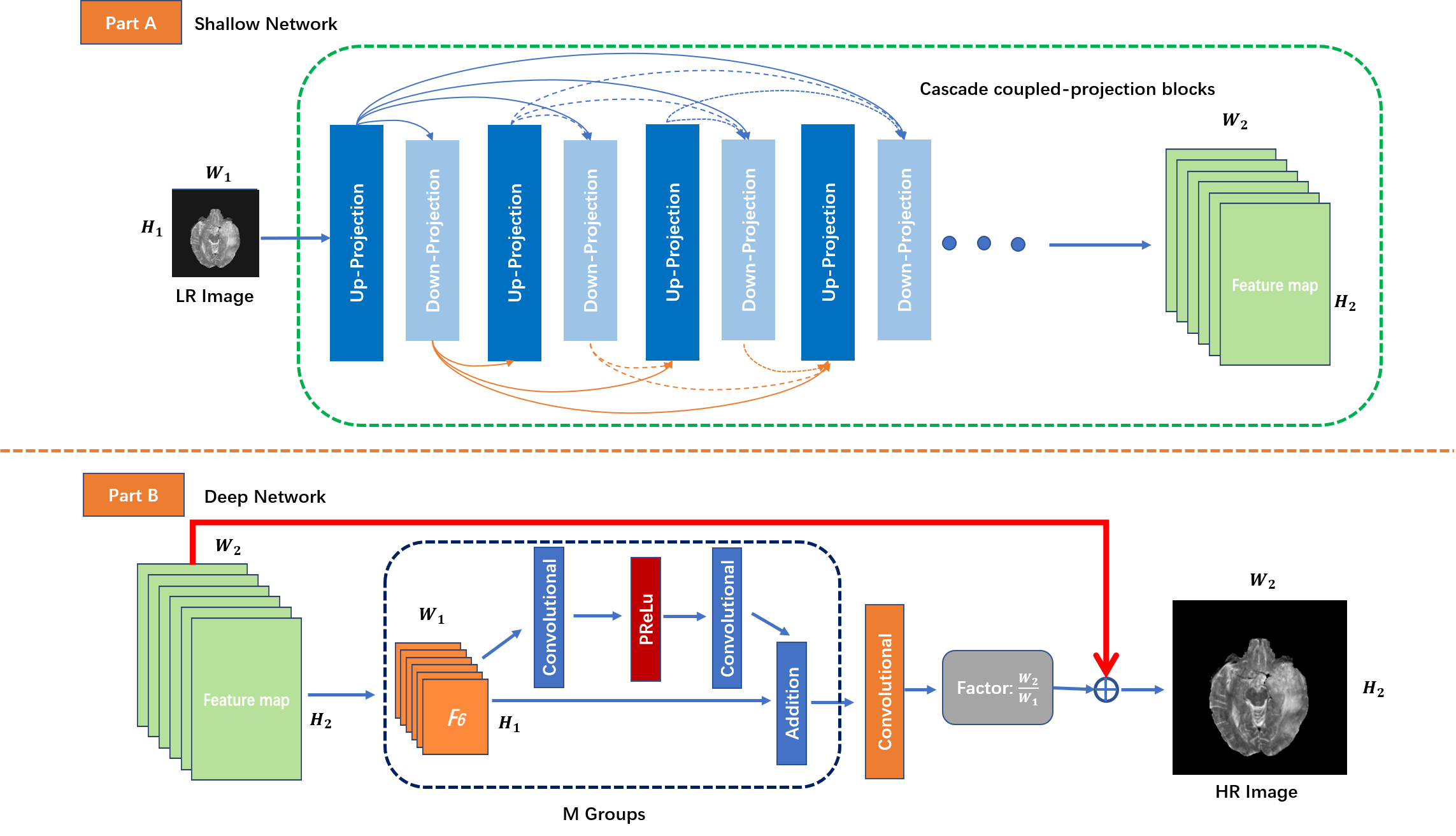}
  \caption{The pipeline of our proposed network. $W$ and $H$ are the dimensions of the feature map. The factor ${W_2}/{W_1}$ is the up-sampling scale.}
  \label{pipeline}
\end{figure*}


Dense-based network feeds the features of each layer in the dense block to all subsequent layers, concatenating the features of all layers instead of adding them directly like residual networks. 
In \cite{40}, Super-Resolution using DenseNet (SRDenseNet) applied dense block structures and improved super-resolution clearly by fusing complementary low-level and high-level features. \cite{61} connects convolutional layers densely through Residual Dense Block (RDB) which better exploits the hierarchical features from all layers. Inspired by the skip-connection, our proposed network constructs step-wise connection to enhance feature propagation in image super-resolution. The step-wise connection from simple to complex is perfectly aligned with the human cognitive process and it also helps flexible embedding in other backbone networks.



The residual-based connection is often too simple and limited without exploiting cross-layer features sufficiently. The dense-based connection improves the cross-layer feature propagation significantly, but it requires a large amount of memory due to channel stacking. Different from the two types of skip-connection approaches, our proposed step-wise connection improves the feature propagation by connecting the corresponding layers of the shallow and deep sub-networks. This design is close to the human from-simple-to-complex cognitive process which helps connect and align cross-layer features effectively. It also reduces memory consumption, more details to be discussed in Experiment part.

%% file: ourwork.tex
In this section, we first present the framework of our method. Then we describe two parts of the proposed CPRN for super-resolution, the deep and shallow network. Finally, we modify and refine our CPRN to strengthen feature propagation.

\subsection{Framework Overview}
Here we present the framework of CPRN in detail. First, in deep learning, we find that shallow networks are good at extracting simple, low-level features (edge features), so as to make the content of reconstructed images consistent with $I^{HR}$ \cite{54,53}. Inspired by Dense DBPN (DDBPN) \cite{48}, a coupled-projection approach is adopted in our shallow network. We build interdependent up- and down-projection blocks, each representing different levels of image degradation and high-resolution components. The shallow network provides a feed-back mechanism for projection errors, so that the network can retain more detail $I^{HR}$ information when generating deeper features. This iterative feedback can help reduce the newly-added noise during the up-sampling stage. Thus, the reconstructed $I^{SR}$ approximates the $I^{HR}$, which is helpful for avoiding consequent misdiagnose due to unreal details generated by noise. Second, since the low-frequency information of $I^{LR}$ and $I^{HR}$ is similar, we only need to obtain the high-frequency difference by the residual network, while inheriting features of shallow networks. A number of residual blocks are cascaded to construct our deep network. The network structure aims to obtain only the residuals so as to ensure good reconstruction. Particularly, the features obtained from the shallow and deep layers are merged, for purpose of getting the final $I^{SR}$. In this way, the $I^{SR}$ could retain more details. 
Third, to contact the shallow and deep network effectively, we refine them with step-wise connection and make it consistent with human cognitive processes (from simple to complex). This step-wise connection network achieves a competitive performance while with fewer convolutional layers, and can be flexibility embedded in other backbone networks.
\begin{algorithm}[tp]
  \caption{ Coupled-projection optimization pseudocode.}
  \label{alg:Framwork}
  \begin{algorithmic}[0]
    \REQUIRE 
      $D_i(x)$: i-th deconvolutional layer, parameterized by $\theta_{d_i}$;\\
      $H_i$: the feature of i-th deconvolutional layer;\\
      $C_i(x)$: i-th convolutional layer, parameterized by $\theta_{c_i}$;\\
      $L_i$: the feature of i-th convolutional layer;
      \STATE \textbf{Input}: Training data $I^{LR}$ (low-resolution images) with $I^{HR}$(high-resolution images);
      \STATE \textbf{Output}: $I^{SR}$ (super-resolution images) and the learnable parameters ($\theta_{d_i}$, $\theta_{c_i}$);
      \STATE \textbf{Initialization}: $\theta_d$, $\theta_c$ are initialized randomly(Gaussian initialization);
    \label{code:fram:extract}
\WHILE{not done}
    \FOR{each epoch}
        \STATE Sample a batch of $I^{LR}_i\sim p(I^{LR})$, $I^{HR}_i\sim q(I^{HR})$,\\ 
        \FOR{all i-th layer}
            \FOR{ all $I^{LR}_i$}
                 \STATE $H_i = D_{i-1}(L_{i-1})$,
                     $L_i = C_i(H_i) $;
                 \STATE ${H_i}^{'}= D_i(|L_i - L_{i-1}|)$, \\
                 $H_{i+1} = H_i^{'} + H_i$.
                     
            \ENDFOR
            \FOR{  all $H_{i+1}$}
                    \STATE $L_{i+1} = C_{i+1}(H_{i+1})$,
                     $H_{i+2} = D_{i+1}(L_{i+1})$;
                    \STATE ${L^{'}_{i+2}} = C_{i+2}(|H_{i+2} - H_{i+1}|)$,\\
                    $L_{i+2} = L^{'}_{i+2} + L_{i+1}$.
            \ENDFOR
        \ENDFOR 
        \STATE Evaluate loss $l^{I^{HR}_i}_{I^{SR}}$ with $I^{SR}$ (the feature of the CPRN) and $HR_i$;
        \STATE \textbf{Update}: $\theta^{'}_d = \theta_d -\alpha\nabla_{\theta_d}l^{I^{HR}_i}_{I^{SR}}$;\\
                     \, \, \, \, \, \, \,            $\theta^{'}c = \theta_c -\alpha\nabla_{\theta_c}l^{I^{HR}_i}_{I^{SR}}$.
    \ENDFOR
\ENDWHILE
  \end{algorithmic}
\end{algorithm}

\begin{figure*}[htp]
\centering
  \includegraphics[width=0.8\textwidth]{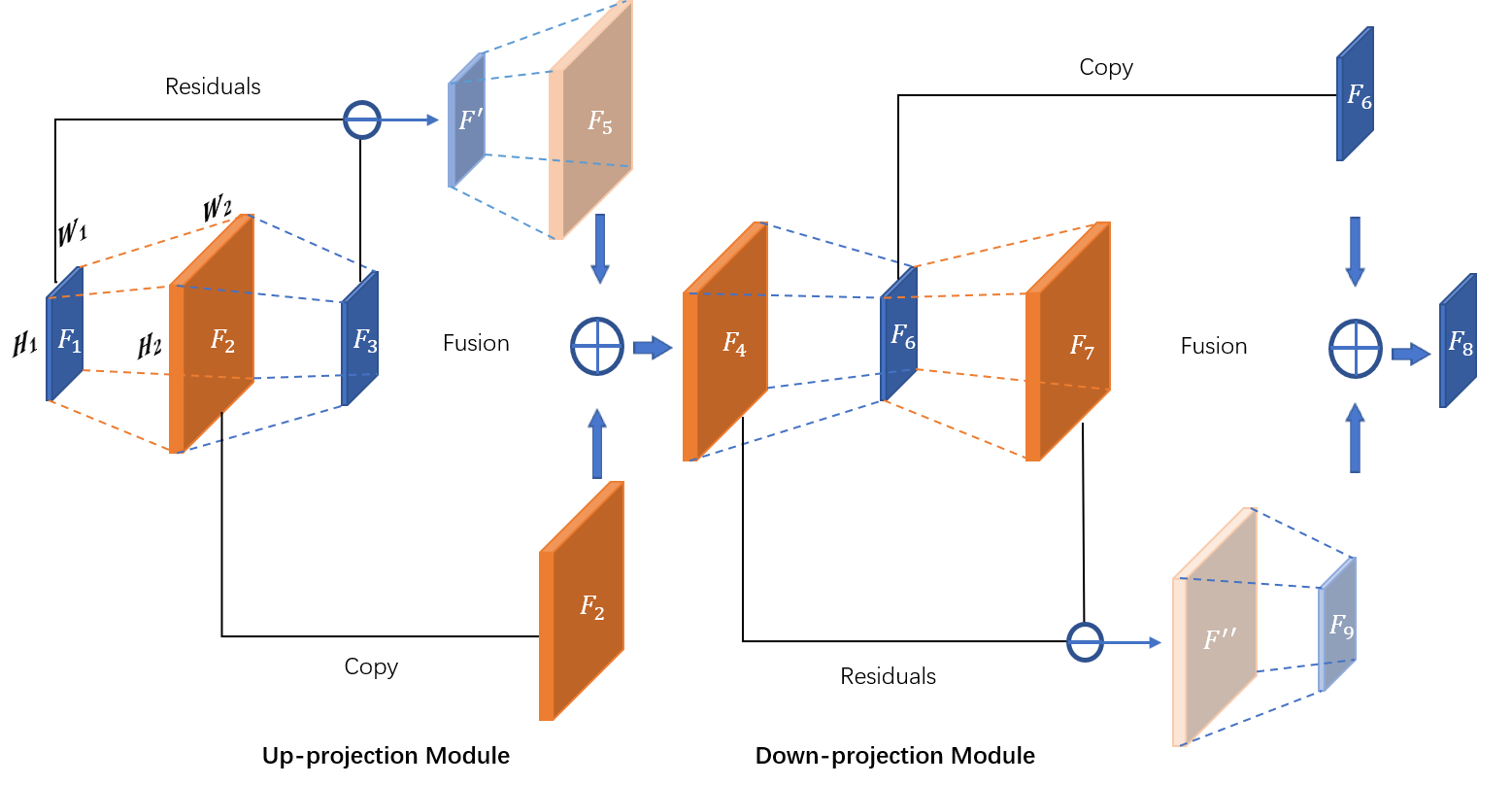}
  \caption{The architecture of coupled-projection block. Up-projection module and down-projection module are presented on left and right side, respectively.}
  \label{coupled_projection}
\end{figure*}

We describe the pipeline of the proposed network CPRN in Fig \ref{pipeline}. The shallow network is illustrated in Part A. We cascade multiple coupled-projection blocks to encourage feature reuse. The coupled-projection is described in Fig. \ref{coupled_projection}, which couples an up-projection module and a down-projection module. Herein $F_i(i=1,…,6)$ is used to denote feature map. $I^{LR}$ images are fed into the convolutional layers to obtain $'F_1'$ with a size of $W_1 \times H_1$. Then, $F_2$ and $F_3$ with different sizes are obtained after a deconvolutional layer and convolutional layer. $F^{'}$ is calculated using the residual errors of $F_1$ and $F_3$, and then send $F^{'}$ into a Deconv(up) layer to obtain $F_5$. We integrate $F_5$ with $F_2$ to get $F_4$. The down-projection block is similar to the up-projection block, except for the following points: 1) the down-projection has two convolutional layers and one deconvolutional layer; 2) the size of $F_5$ is $W_2 \times H_2$ and the size of the final $F_4$ is $W_1 \times H_1$. The deep network is illustrated in Part B. We feed the final feature map from Part A into several convolutional layers to obtain $F_6$. $M$ residual blocks without Batch Normalization (BN) layers are cascaded behind $F_6$. The end of Part B is a convolutional layer with an up-scaling factor $W_2/W_1$, which recovers the super-resolution image. Specially, we utilize the methodology of residual learning to merge the outputs of shallow and deep layers, and then reconstruct the final $I^{SR}$.

\subsection{Shallow Network}


The overall process of the shallow network is described in Algorithm \ref{alg:Framwork}. We use $D_i(x)$ to represent the $i$-th deconvolutional layer, parameterized by $\theta_{di}$, and use $C_i(x)$ to represent the $i$-th convolutional layer, parameterized by $\theta_{ci}$. $L_i$ represents the feature of the $i$-th convolutional layer, and $H_i$ represents the feature of the $i$-th deconvolutional layer. $N$ represents the batch size. We sample low resolution images and high-resolution images from the $I^{LR}$ and $I^{HR}$ datasets, respectively. The previously computed $I^{LR}$ feature map $L_{i-1}$ serves as the input feature. First, the feature map $H_i$ is obtained by mapping $L_{i-1}$ through the deconvolutional layer $D_i$. Then, the feature map $L_i$ is obtained from the convolutional layer $C_i$. Afterwards, the residuals of $L_{i-1}$ and $L_i$ are mapped to the deconvolutional layer so as to get $H_{i}^{'}$. The final $H_{i+1}$ can be obtained by combining $H_{i}^{'}$ with $H_i$. This process is defined as an up-projection block. The down-projection is similar, but the process is in reverse. We firstly map $H_{i+1}$ through the convolutional layer $C_{i+1}$ to obtain the feature map $L_{i+1}$, and then get the feature map $H_{i+1}$ by way of the deconvolutional layer $D_{i+2}$. Afterwards, the $L^{'}_{i+2}$ is obtained by mapping the residuals of $H_{i+2}$ and $H_{i+1}$ to the convolutional layer. Finally, we combine $L^{'}_{i+2}$ and $L_{i+1}$ for the purpose of obtaining $L_{i+1}$. Moreover, to encourage feature reuse and avoid the vanishing-gradient problem, we link the previous up (or down)-projections to the input of each up (or down)-projection. The procedure of shallow network is illustrated in Part A of Fig. \ref{pipeline}.

\subsection{Deep Network}
The deep network of the proposed CPRN architecture can be described as follows:
\begin{align}
F = SR_{s}*Conv(\cdot)\downarrow_s,\\
F_R = R_{es}(\cdot)*(F)\rightarrow_s,\\
I^{SR} = F_{R}*Conv(\cdot)\uparrow_s,
\end{align}
where $SR_s$ is the feature map from the shallow network, and $F$ and $F_R$ are the intermediate feature maps. $s$ is the scaling factor, ($\uparrow$ and $\downarrow$) are the up- and down-sampling operators, respectively, ($\rightarrow$) is the invariant scale; $*$ is the spatial convolution operator; $Conv(\cdot)$ is the convolutional layer and $R_{es}$ is the residual block. First, we send the feature map $SR_s$ obtained from the shallow network to the convolutional layer to obtain the new feature map $F$. Note that the size of $F$ is smaller than $SR_s$. Then, we send $F$ to 16 residual blocks to get $F_R$. The final $I^{SR}$ is obtained by Eq.(3). Note that we reconstruct the final image by merging the $I^{SR}$ features obtained from shallow and deep layers.

The residual network was first proposed to solve a high-level computer vision problem \cite{he2016deep}. Because super-resolution is a low-level computer vision problem, the network structure of ResNet is not completely applicable. To have the network structure satisfy the needs of super-resolution as much as possible, we slim down the network and delete unnecessary modules. The space saved can be used to expand the capacity and depth of the network. In this work, BN layers are removed to prevent the range flexibility of the features from being reduced \cite{haris2018deep}. Because the dropout layer discards many features which might be important for super-resolution, we reconstruct the $I^{LR}$ without incorporating the dropout technique \cite{haris2018deep}.

\subsection{Refinement using Step-wise Connection}

\begin{figure*}[htp]
\centering
  \includegraphics[width=0.9\textwidth]{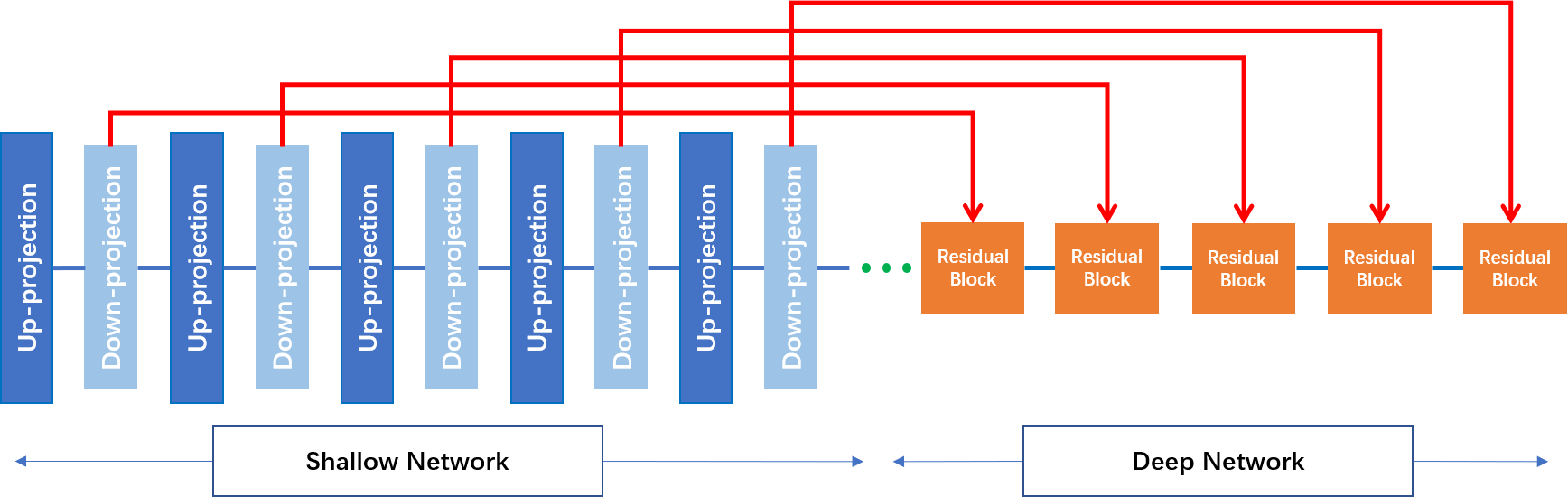}
  \caption{The pipeline of our ste-pwise network CPRN\_S. The feature map which outputted by each down projection in shallow network are fused with the corresponded output of the residual blocks.}
  \label{fig:dense}
\end{figure*}

As the depth of the network increases, the capacitive of the model will gradually increase, and the features will become more abstract. Therefore, it is necessary to contact the shallow and deep network effectively, by which the details in the shallow work can be well preserved. Based on this insight, we refine CPRN with Step-wise connection to make it consistent with human cognitive processes (from simple to complex), named CPRN\_S. The pipeline of our step-wise network CPRN\_S is described in Fig. \ref{fig:dense}. We fuse the feature map which outputted by each down projection in shallow network with the corresponded output of the residual blocks. When the number of residual blocks increases, the incoming features of deep networks become more diverse. Besides, such progressive propagation approach could prevent the input of excessive abstract features and increasing the difficulty of network optimization, when the network capacitive is insufficient. It can be expressed as following:

\begin{equation}
S_{d(i+1)} + D_{Res(i)} = I_{(i+1)},
\end{equation}
where $S_{d(i+1)}$ represents the features from $i$-th down-projection in shallow network, $D_{Res(i)}$ represents the features from $i$-th residual block in shallow network, and $I_{(i+1)}$ represents the input of the $i+1$-th residual block. The improved network not only increases the feature fusion between shallow and deep networks, but also reduces the number of parameters caused by the network depth to a certain extent. And the number of residual block in CPRN\_S is about $1/3$ of the CPRN network. In general, CPRN\_S possesses the advantages of complexity reduction and network distillation.

%% file: experiments.tex
This section presents experimental results including comparisons with the state-of-the-art and detailed analysis of our proposed super-resolution network.

\subsection{Datasets and Evaluation Metrics}
Our proposed super-resolution network CPRN was evaluated over three public brain MRI datasets: Brats, ATLAS\_native, and ATLAS\_standardized. Brats is obtained from the cancer genomics program `The Cancer Genome Atlas' (TCGA, \href{https://www.cancer.gov/about-nci/organization/ccg/research/structural-genomics/tcga}{https://www.cancer.gov/about-nci/organization/ccg/research/structural-genomics/tcga}), which consists of preoperative and multimodal images of glioblastoma and lower grade glioma. The whole dataset contains 102 T1-weighted MRI samples and each sample has 155 image slices. The size of each image is 240$*$240. We read the image from the .nii file of each patient and selected the 60-th, 80-th, 100-th, 120-th and 140-th slices as the $I^{HR}$ data to obtain sufficient brain imaging area. ATLAS\_native and ATLAS\_standardized datasets are obtained from the `Anatomical Tracings of Lesions After Stroke' (ATLAS, \href{https://github.com/npnl/ATLAS/}{https://github.com/npnl/ATLAS/}), which have 304 and 229 T1-weighted MRI samples, respectively. Since the number of image slices in each sample is different, we segmented the image slices of each sample into six groups, and extract an image slice as the $I^{HR}$ from each segmentation point. The $I^{LR}$ is obtained by down-sampling images to $1/2$, $1/4$ of the original resolutions, respectively, via bicubic interpolation. 

Two widely used metrics were utilized in evaluations including peak signal-to-noise ratio (PSNR) and structural similarity (SSIM). Specifically, PSNR evaluates the discrepancy between corresponding pixel points, and SSIM measures image similarity from three aspects including brightness, contrast and structure.

\subsection{Implementation Details}
Our model was implemented on PyTorch 1.0. 
For the standard CPRN, we used 6 coupled-projection blocks and 16 residual blocks. But for the step-wise CPRN\_S, we used 6 coupled-projection blocks and 6 residual blocks. In the shallow sub-network, the kernel sizes were set to 6, 8 on the images of two different sizes, while the strides were set to 2, 4, and padding was set to 2. In the deep sub-network, the kernel size, stride, and padding of the residual blocks were set at 3, 1, and 1, respectively, to keep the size of feature maps unchanged. The number of channels was set at 32 and 64 in the shallow and deep sub-network. During training, the initial learning rate was set at 1e-4 and batch size at 16. The L1 loss function was adopted with Adam where the momentum is 0.9 and the weight decay was 1e-4. All models are trained with 300 epochs and patch size is 48 $\times$ 48. The proposed networks were evaluated over 10, 30, and 30 images from the Brats, ATLAS\_native, and ATLAS\_standardized, respectively.
\begin{table}[h]
\centering
\caption{Super-resolution results of SSIM and PSNR for different methods on MRI dataset. The best results are in red.}
\label{tab:table_1}
\begin{tabular}{@{}cccccc@{}}
\hline
Methods         & Scale & Brats & ATLAS\_native &ATLAS\_standardized\\ \midrule
SRCNN &     $\times$2      &31.423/0.8527      &26.015/0.8737 &23.367/0.8889\\
VDSR &     $\times$2      &34.860/0.9840      &27.270/0.9050 &24.143/0.9148\\
EDSR &     $\times$2      &36.123/0.9027      &27.350/0.9179 &26.042/0.9495\\
DDBPN &     $\times$2      &36.158/0.9867      &28.168/0.9223 &25.671/0.9171\\
CPRN &     $\times$2      &\textcolor[rgb]{1.00,0.00,0.00}{36.500/0.9911}      &\textcolor[rgb]{1.00,0.00,0.00}{28.324/0.9268} &\textcolor[rgb]{1.00,0.00,0.00}{26.114/0.9515}\\ \midrule
SRCNN &     $\times$4      &26.780/0.7034      &20.103/0.7831 &19.998/0.8142\\
VDSR &     $\times$4      &28.739/0.9420      &21.846/0.8087 &20.927/0.8368\\
EDSR &     $\times$4      &29.211/0.7661      &22.217/0.8106 &21.527/0.8578\\
DDBPN &     $\times$4      &29.035/0.9362      &22.207/0.8109 &21.516/0.8595\\
CPRN &     $\times$4      &\textcolor[rgb]{1.00,0.00,0.00}{29.330/0.9439}      &\textcolor[rgb]{1.00,0.00,0.00}{22.331/0.8142} &\textcolor[rgb]{1.00,0.00,0.00}{21.981/0.8687}\\
\hline
\end{tabular}
\end{table}

\begin{figure*}[htp]
\centering
	\includegraphics[width=1.0\textwidth]{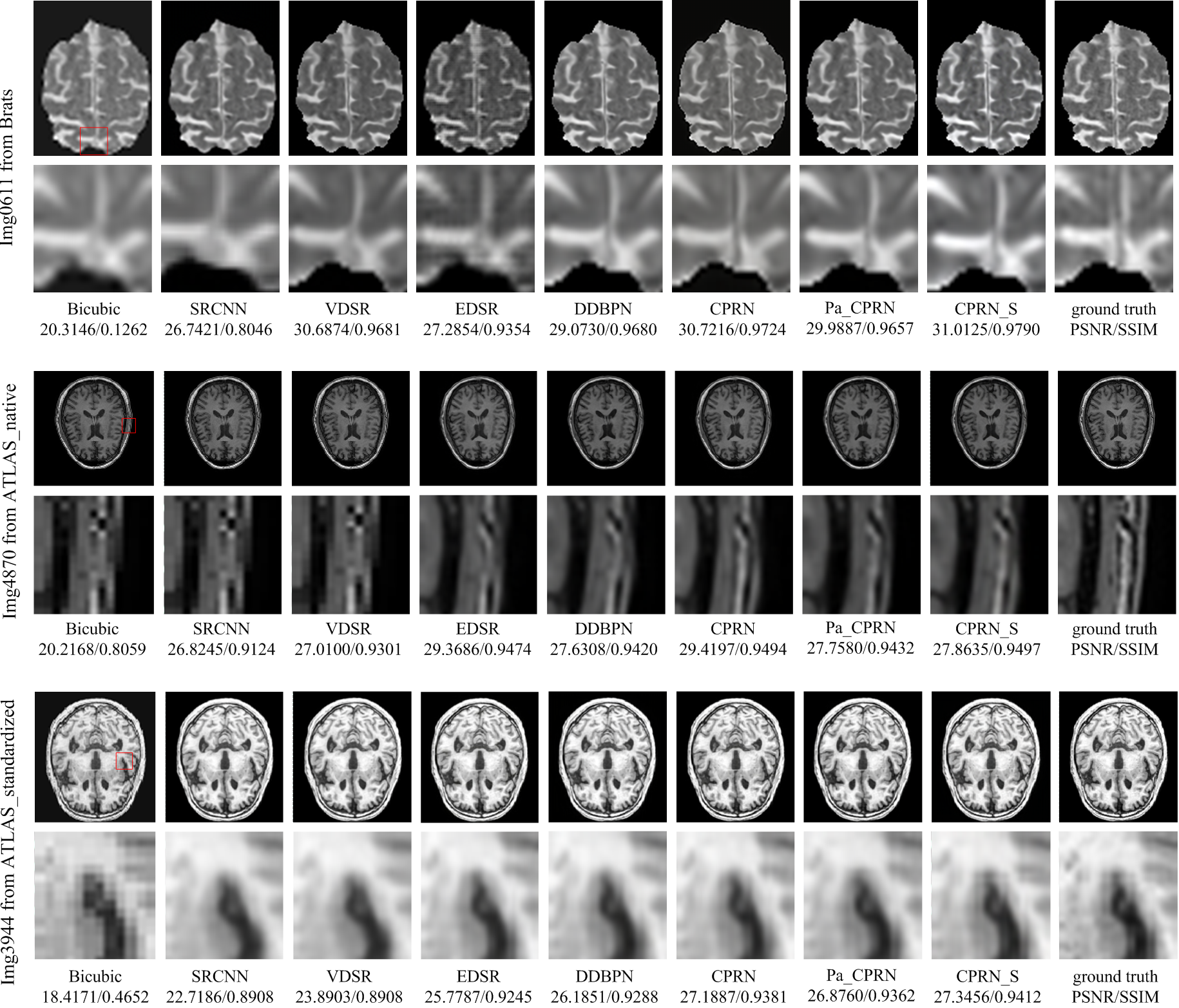}
	\caption{Graphical representation of qualitative comparison in the $\times$2 enlargement between our models and baseline methods on Brats dataset. Pa\_CPRN is the ablation model which shallow and deep networks are executed in parallel. CPRN\_S is the enhanced version with step-wise connection.}
	\label{fig:result}
\end{figure*}


\subsection{Comparison with baseline methods}
We compare our networks with several state-of-the-art networks SRCNN \cite{1}, VDSR \cite{43}, EDSR \cite{39}, and DDBPN \cite{48}, and the comparison is based on two up-scaling by $\times2$ and $\times4$. Table \ref{tab:table_1} shows the super-resolution ($\times2$ and $\times4$) for the dataset Brats, ATLAS\_native and ATLAS\_standardized dataset. As Table \ref{tab:table_1} shows, our networks achieve better PSNR and SSIM consistently under both up-scaling cases, and this largely attribute to the complementary shallow and deep sub-networks that help retain fine-detailed features in the high-resolution images. Specifically, CPRN achieves comparable PSNR and SSIM with other monotonous models, revealing the robustness of our two-stage network in medical image super-resolution. 
VDSR and SRCNN obtain relatively poorer performance, largely due to their redundant structures where some unnecessary network layers such as BN could reduce the range flexibility of the features. In addition, VDSR, EDSR, and DDBPN show lower consistency across the three datasets, e.g. DDBPN performs better than EDSR on Brats and ATLAS\_native but worse than EDSR on ATLAS\_standardized for the $\times$2 up-scaling.

Fig. \ref{fig:result} shows the super-resolution images by our network and the compared ones over the three dataset. As Fig. \ref{fig:result} show, the super-resolution images are well aligned with the quantitative PSNR and SSIM where our CPRN produces clearer image reconstruction with better details. Specifically, the blood vessels in $I^{SR}$ by our network retain higher consistency with the ground-truth $I^{HR}$ to a large extent. The boundary between the blood vessels and gray matter is clearer with high similarity to the ground-truth $I^{HR}$. All these show that our CPRN preserves more high-resolution components than other networks and reconstructs quality image with more detailed features. As a comparison, SRCNN, VDSR, EDSR, and DDBPN tend to generate misleading information in several cases.
Specifically, the EDSR produces a stripe pattern in its reconstructed images over the Brats dataset, and its reconstructed images over the other two datasets are also vague and blurry. The Pa\_CPRN and CPRN\_S are two CPRN variants which will be discussed in Ablation study in the ensuing subsection.



\begin{table}[h]
\centering
\caption{Super-resolution results of SSIM and PSNR for different methods on MRI dataset. The best results are in red, the second best results are in blue.}
\setlength{\tabcolsep}{1.5mm}{
\begin{tabular}{@{}ccccc@{}}
\hline
Methods         & Scale & Brats & ATLAS\_native &ATLAS\_standardized\\ \midrule
CP\_SD & $\times$2 &36.439/0.9904 &28.048/\textcolor[rgb]{0.00,0.00,1.00}{0.9213} &25.977/0.9487  \\
RN\_SD & $\times$2 &36.357/0.9882 &27.600/0.9158 &25.989/0.9420  \\
Pa\_CPRN &     $\times$2      &36.346/0.9856      &28.292/0.9166 &25.996/\textcolor[rgb]{0.00,0.00,1.00}{0.9501}\\
CPRN &     $\times$2      &\textcolor[rgb]{0.00,0.00,1.00}{36.500/0.9911}      &\textcolor[rgb]{0.00,0.00,1.00}{28.324}/\textcolor[rgb]{1.00,0.00,0.00}{0.9268} &\textcolor[rgb]{0.00,0.00,1.00}{26.114}/\textcolor[rgb]{1.00,0.00,0.00}{0.9515}\\
CPRN\_S &     $\times$2      &\textcolor[rgb]{1.00,0.00,0.00}{36.543/0.9918}      &\textcolor[rgb]{1.00,0.00,0.00}{28.512}/0.9187 &\textcolor[rgb]{1.00,0.00,0.00}{26.673}/0.9345\\
\midrule
CP\_SD & $\times$4 &28.995/0.9356 &\textcolor[rgb]{0.00,0.00,1.00}{22.389}/0.8103 &21.925/0.8654 \\
RN\_SD & $\times$4 &28.947/0.7865 &22.326/0.8131 &21.912/0.8651 \\
Pa\_CPRN &     $\times$4      &29.174/0.9348      &22.359/0.8128 &21.598/0.8604\\
CPRN &     $\times$4      &\textcolor[rgb]{0.00,0.00,1.00}{29.330/0.9439}      &22.331/\textcolor[rgb]{0.00,0.00,1.00}{0.8142} &\textcolor[rgb]{0.00,0.00,1.00}{21.981/0.8687}\\
CPRN\_S &     $\times$4      &\textcolor[rgb]{1.00,0.00,0.00}{29.445/0.9501}      &\textcolor[rgb]{1.00,0.00,0.00}{22.701/0.8175} &\textcolor[rgb]{1.00,0.00,0.00}{22.258/0.8809}\\

\hline
\end{tabular}
\label{tab:table_2}}
\end{table}

\subsection{Ablation study}
Our proposed CPRN consists of a deep sub-network and a shallow sub-network as well as a step-wise connection mechanism for better sup-resolution performance. We evaluated three network architectures beyond the standard CPRN to investigate how these designs contribute to the overall performance: 1) A monotonous version that executes either the coupled-projection blocks (CP\_SD) or the residual blocks (RN\_SD) alone; 2) A parallel version Pa\_CPRN that executes shallow and deep sub-networks in parallel and averages their outputs as the final output; 3) A step-wise version CPRN\_S that improves CPRN with our proposed step-wise connection mechanism. Table \ref{tab:table_2} shows experimental results. As Table \ref{tab:table_2} shows, CP\_SD and RN\_SD obtain lower PSNR and SSIM as they only process either coupled-projection blocks or residual blocks alone without fusing features of different types. Pa\_CPRN combines the two sub-network in parallel which does not show clear performance improvement, showing that running the two sub-networks in parallel does not capture their merits. CPRN\_S and CPRN obtain the best performance on Brats and ATLAS\_standardized datasets, of which CPRN\_S performs better than CPRN especially on the $\times4$ scenario. Additionally, CPRN\_S uses much less parameters than CPRN, more details to be discussed in the following Discussion part.

\begin{figure*}[htp]
\centering
	\includegraphics[width=0.9\textwidth]{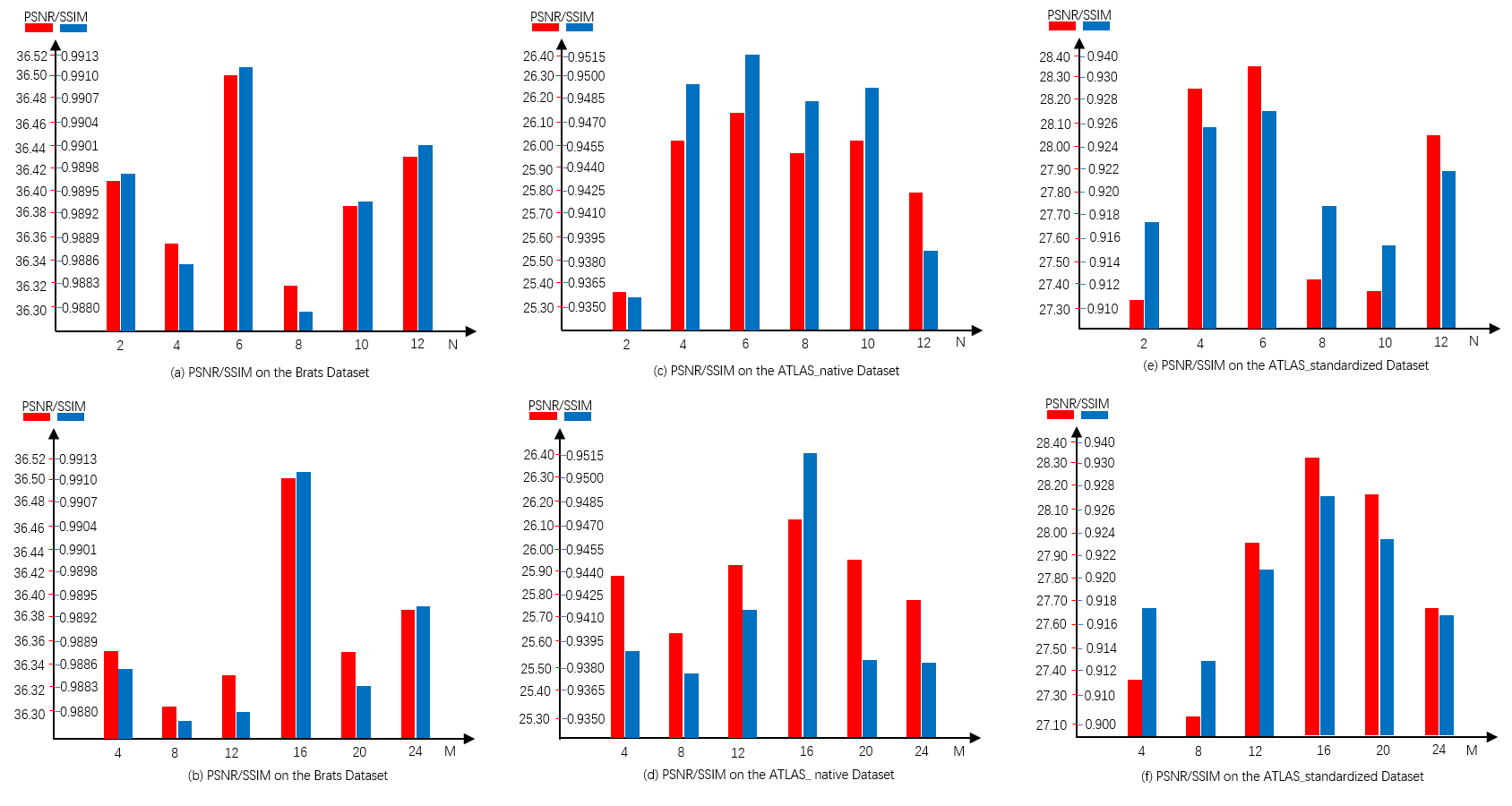}
	\caption{Results of PSNR and SSIM under different $N$ and $M$ values. The network performs best when $N = 6$ and $M = 16$.}
	\label{fig:buchongshiyan}
\end{figure*}

Fig. \ref{fig:result} illustrates the super-resolution by the CPRN, Pa\_CPRN, and CPRN\_S, respectively. As Fig. \ref{fig:result} shows, CPRN and CPRN\_S generate clearer texture patterns than Pa\_CPRN, and the reconstructed images are subjectively closer to the ground truth. This shows that the proposed two-stage network structure in CPRN can retain more image details and reconstruct clearer images. CPRN\_S propagates features progressively by step-wise connecting the output of each down projection in the shallow sub-network with the corresponding output of the residual block in the deep sub-network. Such step-wise connection mechanism is similar to the human cognitive process which helps the network to achieve great super-resolution performance.

\begin{table}[t]
\centering
\caption{Results of different numbers of channels in the deep and shallow networks. $sc$ is the number of channels in the shallow network, and $dn$ is the number of channels in the deep network. The best results are in red.}
\begin{tabular}{@{}cccccc@{}}
\hline
Channels                     & Brats & ATLAS\_native &ATLAS\_standardized\\ \midrule
$sc = 64, dc = 64$           &36.425/0.9887      &27.877/0.9184 &25.457/0.9402\\
$sc = 16, dc = 64$           &36.320/0.9821      &27.232/0.9172 &25.428/0.9412\\
$sc = 32, dc = 32$           &36.382/0.9834      &27.441/0.9198 &25.811/0.9453\\
$sc = 32, dc = 128$          &29.330/0.9439      &22.331/0.8142 &21.981/0.8687\\
$sc = 32, dc = 64$           &\textcolor[rgb]{1.00,0.00,0.00}{36.500/0.9911}      &\textcolor[rgb]{1.00,0.00,0.00}{28.324/0.9268} &\textcolor[rgb]{1.00,0.00,0.00}{26.114/0.9515}\\
\hline
\end{tabular}
\label{tab:table_3}
\end{table}

\begin{table}[t]
\centering
\caption{Results of BN layers in shallow and deep network. Here ($*$,$*$) represents whether or not BN layers are used in the shallow and deep networks. wBN and w$/$oBN represent using and not using BN layers, respectively. The best results are in red.}
\begin{tabular}{@{}cccccc@{}}
\hline
Channels                     & Brats & ATLAS\_native &ATLAS\_standardized\\ \midrule
wBN, w/oBN           &36.312/0.6420      &27.338/0.9089 &25.777/0.9327\\
wBN, wBN          &36.484/0.9632      &27.262/0.9104 &25.655/0.9421\\
w/oBN, wBN           &36.153/0.8668      &27.163/0.8682 &25.621/0.8924\\
w/oBN, w/oBN          &\textcolor[rgb]{1.00,0.00,0.00}{36.500/0.9911}      &\textcolor[rgb]{1.00,0.00,0.00}{28.324/0.9268} &\textcolor[rgb]{1.00,0.00,0.00}{26.114/0.9515}\\
\hline
\end{tabular}
\label{tab:table_4}
\end{table}

\begin{table}[t]
\centering
\caption{Computation time of each method on three datasets.}
\begin{tabular}{@{}cccccc@{}}
\hline
Methods         & Brats & ATLAS\_native &ATLAS\_standardized\\ \midrule
SRCNN           &1.76(s)      &1.83(s) &1.74(s)\\
VDSR           &3.45(s)      &3.64(s) &3.27(s)\\
EDSR           &2.37(s)      &2.52(s) &3.03(s)\\
DDBPN          &2.58(s)      &2.91(s) &2.36(s)\\
CPRN          &4.36(s)      &4.25(s) &4.67(s)\\
Pa\_CPRN          &4.07(s)      &4.21(s) &4.23(s)\\
CPRN\_S          &3.48(s)      &3.56(s) &3.52(s)\\
CP\_S          &8.48(s)      &8.59(s) &8.42(s)\\
RN\_S          &3.37(s)      &3.49(s) &3.78(s)\\
\hline
\end{tabular}
\label{tab:table_5}
\end{table}

\subsection{Discussion}
To further verify the effectiveness of our proposed network structure, we experimented using different $N$ and $M$ and Fig. \ref{fig:buchongshiyan} shows the obtained PSNR and SSIM. Specifically, we first fixed $M$ to measure the effect of different $N$. The results show that PSNR and SSIM perform the best when $N = 6$. In addition, we fixed $N$ to measure the effect of different $M$, and found that the network obtains the best performance when $M = 16$. Further, the two parameters $N$ and $M$ are quite consistent which achieve the optimal performance when they are set around 6 and 16 across the three studied datasets.

\begin{figure}[tb]
\centering
	\includegraphics[width=0.48\textwidth]{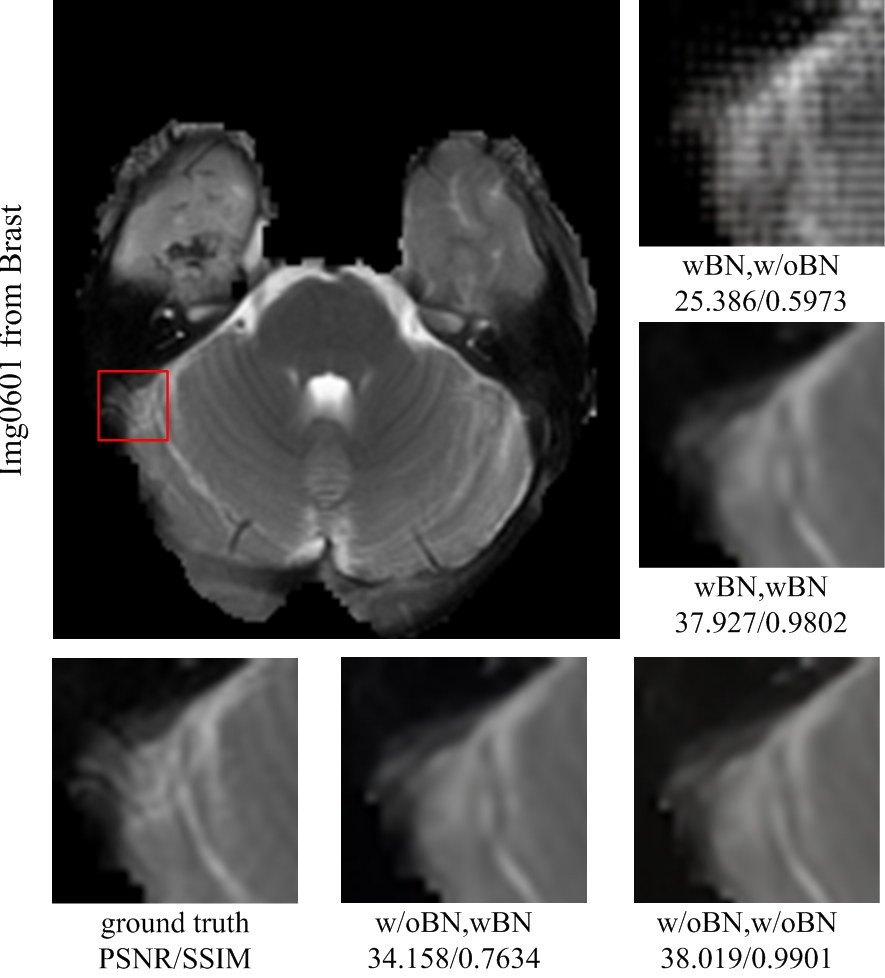}
	\caption{Graphical representation of BN layers in shallow and deep network. These figures show that removing the BN layers both in shallow and deep networks can better reconstruct texture details.}
	\label{fig:bn_results}
\end{figure}

The numbers of channels in the deep sub-network ($dc$) and shallow sub-network ($sc$) also affect the image super-resolution performance. Table \ref{tab:table_3} shows experimental results. As Table \ref{tab:table_3} shows, the proposed CPRN achieve the optimal performance when $sc$ and $dc$ are set around 32 and 64, respectively, consistently across the three studied datasets. Specifically, the PSNR and SSIM drop sharply when $dc$ becomes larger. In addition, increasing $sc$ in the shallow sub-network empowers the network to better learn the relationship between $I^{SR}$ and $I^{HR}$ during the coupled-projection, and the reconstructed features were consistent with that of $I^{HR}$. On the other hand, the relationship between $I^{SR}$ and $I^{HR}$ features was not learned well when $sc$ was too small.

Previous works show difference influence of BN layer while introduced for image super-resolution \cite{62,48,39}. We also studied the effect of introducing BN layers in our shallow and deep sub-networks. Table \ref{tab:table_4} shows experimental results. As Table \ref{tab:table_4} shows, our proposed CPRN achieves the highest PSNR and SSIM without using the BN layers. Fig. \ref{fig:bn_results} further illustrates that removing BN layers in the shallow and deep sub-networks can better reconstruct texture details. The three cases on the right tend to produce stripe patterns or blurred details that degrade the image super-resolution performance. This study show that it is good to exclude BN layers which helps to get better super-resolution images and also reduce the network parameters and complexity.

We also studied the computational cost of different super-resolution networks, and explored how different network structures influence the speed of our proposed models (CPRN, Pa\_CPRN, CPRN\_S). Table \ref{tab:table_5} shows the experimental results where the execution time of each network is evaluated over 30 dataset images. It can be seen that SRCNN takes the shortest time as it has only three fully connected layers. VDSR takes more time than EDSR and DDBPN as VDSR has a much deep structure by cascading 20 residual blocks with BN layers. Although the depth of Pa\_CPRN is reduced, the increase of the network width leads to an increase in the number of channels. CPRN\_S use step-wise connection to progressively and hierarchically spread and reuse features and this design reduces the network depth and meanwhile avoids increasing the number of channels. Its computational cost is therefore lower than that of Pa\_CPRN and CPRN. Specifically, the parameters of CPRN in shallow network are similar to the deep network and CPRN\_S saves nearly 60\% of the parameters compared to CPRN, CPRN\_S therefore saves around 30\% of network parameters as compared with CPRN, leading to the lower computational costs.

%% file: conclusion.tex
This paper presented a deep network for MRI super-resolution. Unlike the previous methods whose network structures are monotonous, our proposed network consists of two main parts: a shallow network and a deep one. In the shallow network, a coupled-projection mechanism helps to build interconnected up- and down-projection stages, making the reconstructed $I^{SR}$ closer to the $I^{HR}$ in content. In the deep network, we cascade multiple residual blocks to attain the high-frequency residuals of the $I^{LR}$ and $I^{HR}$. Furthermore, to be consistent with human cognitive processes (from simple to complex), we develop an enhanced version CPRN\_S which stepwisely connects each down projection in the shallow network with the corresponded residual blocks. The results in terms of PSNR and SSIM, show that our network significantly outperforms current state-of-the-art methods in reconstructing clear images from degraded ones.

%% file: bare_jrnl.bbl
\begin{thebibliography}{10}
\providecommand{\url}[1]{#1}
\csname url@samestyle\endcsname
\providecommand{\newblock}{\relax}
\providecommand{\bibinfo}[2]{#2}
\providecommand{\BIBentrySTDinterwordspacing}{\spaceskip=0pt\relax}
\providecommand{\BIBentryALTinterwordstretchfactor}{4}
\providecommand{\BIBentryALTinterwordspacing}{\spaceskip=\fontdimen2\font plus
\BIBentryALTinterwordstretchfactor\fontdimen3\font minus
  \fontdimen4\font\relax}
\providecommand{\BIBforeignlanguage}[2]{{%
\expandafter\ifx\csname l@#1\endcsname\relax
\typeout{** WARNING: IEEEtran.bst: No hyphenation pattern has been}%
\typeout{** loaded for the language `#1'. Using the pattern for}%
\typeout{** the default language instead.}%
\else
\language=\csname l@#1\endcsname
\fi
#2}}
\providecommand{\BIBdecl}{\relax}
\BIBdecl

\bibitem{3}
K.~Christensen-Jeffries, R.~J. Browning, M.-X. Tang, C.~Dunsby, and R.~J.
  Eckersley, ``In vivo acoustic super-resolution and super-resolved velocity
  mapping using microbubbles,'' \emph{IEEE transactions on medical imaging},
  vol.~34, no.~2, pp. 433--440, 2014.

\bibitem{2}
C.~Cruz, R.~Mehta, V.~Katkovnik, and K.~O. Egiazarian, ``Single image
  super-resolution based on wiener filter in similarity domain,'' \emph{IEEE
  Transactions on Image Processing}, vol.~27, no.~3, pp. 1376--1389, 2017.

\bibitem{1}
C.~Dong, C.~C. Loy, K.~He, and X.~Tang, ``Image super-resolution using deep
  convolutional networks,'' \emph{IEEE transactions on pattern analysis and
  machine intelligence}, vol.~38, no.~2, pp. 295--307, 2015.

\bibitem{tip_4}
Y.~Huang, L.~Shao, and A.~F. Frangi, ``Simultaneous super-resolution and
  cross-modality synthesis of 3d medical images using weakly-supervised joint
  convolutional sparse coding,'' in \emph{Proceedings of the IEEE Conference on
  Computer Vision and Pattern Recognition}, 2017, pp. 6070--6079.

\bibitem{7}
T.~Uiboupin, P.~Rasti, G.~Anbarjafari, and H.~Demirel, ``Facial image super
  resolution using sparse representation for improving face recognition in
  surveillance monitoring,'' in \emph{2016 24th Signal Processing and
  Communication Application Conference (SIU)}.\hskip 1em plus 0.5em minus
  0.4em\relax IEEE, 2016, pp. 437--440.

\bibitem{6}
P.~Rasti, T.~Uiboupin, S.~Escalera, and G.~Anbarjafari, ``Convolutional neural
  network super resolution for face recognition in surveillance monitoring,''
  in \emph{International conference on articulated motion and deformable
  objects}.\hskip 1em plus 0.5em minus 0.4em\relax Springer, 2016, pp.
  175--184.

\bibitem{8}
R.~Salman and I.~Willms, ``A mobile security robot equipped with uwb-radar for
  super-resolution indoor positioning and localisation applications,'' in
  \emph{2012 International Conference on Indoor Positioning and Indoor
  Navigation (IPIN)}.\hskip 1em plus 0.5em minus 0.4em\relax IEEE, 2012, pp.
  1--8.

\bibitem{10}
M.~Filippi, M.~A. Rocca, O.~Ciccarelli, N.~De~Stefano, N.~Evangelou, L.~Kappos,
  A.~Rovira, J.~Sastre-Garriga, M.~Tintor{\`e}, J.~L. Frederiksen
  \emph{et~al.}, ``Mri criteria for the diagnosis of multiple sclerosis:
  Magnims consensus guidelines,'' \emph{The Lancet Neurology}, vol.~15, no.~3,
  pp. 292--303, 2016.

\bibitem{9}
S.~Pereira, A.~Pinto, V.~Alves, and C.~A. Silva, ``Brain tumor segmentation
  using convolutional neural networks in mri images,'' \emph{IEEE transactions
  on medical imaging}, vol.~35, no.~5, pp. 1240--1251, 2016.

\bibitem{11}
D.~Owen, A.~Melbourne, Z.~Eaton-Rosen, D.~L. Thomas, N.~Marlow, J.~Rohrer, and
  S.~Ourselin, ``Deep convolutional filtering for spatio-temporal denoising and
  artifact removal in arterial spin labelling mri,'' in \emph{International
  Conference on Medical Image Computing and Computer-Assisted
  Intervention}.\hskip 1em plus 0.5em minus 0.4em\relax Springer, 2018, pp.
  21--29.

\bibitem{12}
C.~Andersson, J.~Kihlberg, T.~Ebbers, L.~Lindstr{\"o}m, C.-J. Carlh{\"a}ll, and
  J.~E. Engvall, ``Phase-contrast mri volume flow--a comparison of breath held
  and navigator based acquisitions,'' \emph{BMC medical imaging}, vol.~16,
  no.~1, p.~26, 2016.

\bibitem{13}
N.~Zhang, G.~Yang, Z.~Gao, C.~Xu, Y.~Zhang, R.~Shi, J.~Keegan, L.~Xu, H.~Zhang,
  Z.~Fan \emph{et~al.}, ``Deep learning for diagnosis of chronic myocardial
  infarction on nonenhanced cardiac cine mri,'' \emph{Radiology}, p. 182304,
  2019.

\bibitem{15}
N.~Basty and V.~Grau, ``Super resolution of cardiac cine mri sequences using
  deep learning,'' in \emph{Image Analysis for Moving Organ, Breast, and
  Thoracic Images}.\hskip 1em plus 0.5em minus 0.4em\relax Springer, 2018, pp.
  23--31.

\bibitem{14}
Y.~Chen, Y.~Xie, Z.~Zhou, F.~Shi, A.~G. Christodoulou, and D.~Li, ``Brain mri
  super resolution using 3d deep densely connected neural networks,'' in
  \emph{2018 IEEE 15th International Symposium on Biomedical Imaging (ISBI
  2018)}.\hskip 1em plus 0.5em minus 0.4em\relax IEEE, 2018, pp. 739--742.

\bibitem{5}
T.~K{\"o}hler, ``Multi-frame super-resolution reconstruction with applications
  to medical imaging,'' \emph{arXiv preprint arXiv:1812.09375}, 2018.

\bibitem{18}
Y.~Zhang, Y.~Zhang, W.~Li, Y.~Huang, and J.~Yang, ``Super-resolution surface
  mapping for scanning radar: Inverse filtering based on the fast iterative
  adaptive approach,'' \emph{IEEE Transactions on Geoscience and Remote
  Sensing}, vol.~56, no.~1, pp. 127--144, 2017.

\bibitem{19}
W.~Dong, F.~Fu, G.~Shi, X.~Cao, J.~Wu, G.~Li, and X.~Li, ``Hyperspectral image
  super-resolution via non-negative structured sparse representation,''
  \emph{IEEE Transactions on Image Processing}, vol.~25, no.~5, pp. 2337--2352,
  2016.

\bibitem{20}
J.~Yang, J.~Wright, T.~S. Huang, and Y.~Ma, ``Image super-resolution via sparse
  representation,'' \emph{IEEE transactions on image processing}, vol.~19,
  no.~11, pp. 2861--2873, 2010.

\bibitem{21}
M.~Irani and S.~Peleg, ``Improving resolution by image registration,''
  \emph{CVGIP: Graphical models and image processing}, vol.~53, no.~3, pp.
  231--239, 1991.

\bibitem{22}
G.~Freedman and R.~Fattal, ``Image and video upscaling from local
  self-examples,'' \emph{ACM Transactions on Graphics (TOG)}, vol.~30, no.~2,
  p.~12, 2011.

\bibitem{23}
J.~Yang, J.~Wright, T.~Huang, and Y.~Ma, ``Image super-resolution as sparse
  representation of raw image patches,'' in \emph{2008 IEEE conference on
  computer vision and pattern recognition}.\hskip 1em plus 0.5em minus
  0.4em\relax Citeseer, 2008, pp. 1--8.

\bibitem{24}
J.~Xie, R.~S. Feris, and M.-T. Sun, ``Edge-guided single depth image super
  resolution,'' \emph{IEEE Transactions on Image Processing}, vol.~25, no.~1,
  pp. 428--438, 2015.

\bibitem{27}
W.~Shi, J.~Caballero, F.~Husz{\'a}r, J.~Totz, A.~P. Aitken, R.~Bishop,
  D.~Rueckert, and Z.~Wang, ``Real-time single image and video super-resolution
  using an efficient sub-pixel convolutional neural network,'' in
  \emph{Proceedings of the IEEE conference on computer vision and pattern
  recognition}, 2016, pp. 1874--1883.

\bibitem{26}
C.~Dong, C.~C. Loy, and X.~Tang, ``Accelerating the super-resolution
  convolutional neural network,'' in \emph{European conference on computer
  vision}.\hskip 1em plus 0.5em minus 0.4em\relax Springer, 2016, pp. 391--407.

\bibitem{25}
C.~Dong, C.~C. Loy, K.~He, and X.~Tang, ``Learning a deep convolutional network
  for image super-resolution,'' in \emph{European conference on computer
  vision}.\hskip 1em plus 0.5em minus 0.4em\relax Springer, 2014, pp. 184--199.

\bibitem{16}
Y.~Chen, F.~Shi, A.~G. Christodoulou, Y.~Xie, Z.~Zhou, and D.~Li, ``Efficient
  and accurate mri super-resolution using a generative adversarial network and
  3d multi-level densely connected network,'' in \emph{International Conference
  on Medical Image Computing and Computer-Assisted Intervention}.\hskip 1em
  plus 0.5em minus 0.4em\relax Springer, 2018, pp. 91--99.

\bibitem{32}
L.~Zhang and X.~Wu, ``An edge-guided image interpolation algorithm via
  directional filtering and data fusion,'' \emph{IEEE transactions on Image
  Processing}, vol.~15, no.~8, pp. 2226--2238, 2006.

\bibitem{34}
Y.-W. Tai, S.~Liu, M.~S. Brown, and S.~Lin, ``Super resolution using edge prior
  and single image detail synthesis,'' in \emph{2010 IEEE Computer Society
  Conference on Computer Vision and Pattern Recognition}.\hskip 1em plus 0.5em
  minus 0.4em\relax IEEE, 2010, pp. 2400--2407.

\bibitem{33}
J.~Sun, Z.~Xu, and H.-Y. Shum, ``Image super-resolution using gradient profile
  prior,'' in \emph{2008 IEEE Conference on Computer Vision and Pattern
  Recognition}.\hskip 1em plus 0.5em minus 0.4em\relax IEEE, 2008, pp. 1--8.

\bibitem{35}
L.~Zhang, W.~Wei, C.~Bai, Y.~Gao, and Y.~Zhang, ``Exploiting clustering
  manifold structure for hyperspectral imagery super-resolution,'' \emph{IEEE
  Transactions on Image Processing}, vol.~27, no.~12, pp. 5969--5982, 2018.

\bibitem{36}
R.~Zeyde, M.~Elad, and M.~Protter, ``On single image scale-up using
  sparse-representations,'' in \emph{International conference on curves and
  surfaces}.\hskip 1em plus 0.5em minus 0.4em\relax Springer, 2010, pp.
  711--730.

\bibitem{37}
R.~Timofte, V.~De~Smet, and L.~Van~Gool, ``A+: Adjusted anchored neighborhood
  regression for fast super-resolution,'' in \emph{Asian conference on computer
  vision}.\hskip 1em plus 0.5em minus 0.4em\relax Springer, 2014, pp. 111--126.

\bibitem{38}
X.~Gao, K.~Zhang, D.~Tao, and X.~Li, ``Image super-resolution with sparse
  neighbor embedding,'' \emph{IEEE Transactions on Image Processing}, vol.~21,
  no.~7, pp. 3194--3205, 2012.

\bibitem{40}
T.~Tong, G.~Li, X.~Liu, and Q.~Gao, ``Image super-resolution using dense skip
  connections,'' in \emph{Proceedings of the IEEE International Conference on
  Computer Vision}, 2017, pp. 4799--4807.

\bibitem{39}
B.~Lim, S.~Son, H.~Kim, S.~Nah, and K.~Mu~Lee, ``Enhanced deep residual
  networks for single image super-resolution,'' in \emph{Proceedings of the
  IEEE Conference on Computer Vision and Pattern Recognition Workshops}, 2017,
  pp. 136--144.

\bibitem{30}
C.~Ledig, L.~Theis, F.~Husz{\'a}r, J.~Caballero, A.~Cunningham, A.~Acosta,
  A.~Aitken, A.~Tejani, J.~Totz, Z.~Wang \emph{et~al.}, ``Photo-realistic
  single image super-resolution using a generative adversarial network,'' in
  \emph{Proceedings of the IEEE conference on computer vision and pattern
  recognition}, 2017, pp. 4681--4690.

\bibitem{50}
O.~Oktay, W.~Bai, M.~Lee, R.~Guerrero, K.~Kamnitsas, J.~Caballero,
  A.~de~Marvao, S.~Cook, D.~O’Regan, and D.~Rueckert, ``Multi-input cardiac
  image super-resolution using convolutional neural networks,'' in
  \emph{International conference on medical image computing and
  computer-assisted intervention}.\hskip 1em plus 0.5em minus 0.4em\relax
  Springer, 2016, pp. 246--254.

\bibitem{28}
C.-H. Pham, A.~Ducournau, R.~Fablet, and F.~Rousseau, ``Brain mri
  super-resolution using deep 3d convolutional networks,'' in \emph{2017 IEEE
  14th International Symposium on Biomedical Imaging (ISBI 2017)}.\hskip 1em
  plus 0.5em minus 0.4em\relax IEEE, 2017, pp. 197--200.

\bibitem{29}
C.~Zhao, A.~Carass, B.~E. Dewey, J.~Woo, J.~Oh, P.~A. Calabresi, D.~S. Reich,
  P.~Sati, D.~L. Pham, and J.~L. Prince, ``A deep learning based anti-aliasing
  self super-resolution algorithm for mri,'' in \emph{International Conference
  on Medical Image Computing and Computer-Assisted Intervention}.\hskip 1em
  plus 0.5em minus 0.4em\relax Springer, 2018, pp. 100--108.

\bibitem{31}
A.~Lucas, S.~Lopez-Tapiad, R.~Molinae, and A.~K. Katsaggelos, ``Generative
  adversarial networks and perceptual losses for video super-resolution,''
  \emph{IEEE Transactions on Image Processing}, 2019.

\bibitem{51}
L.~Han and Z.~Yin, ``A cascaded refinement gan for phase contrast microscopy
  image super resolution,'' in \emph{International Conference on Medical Image
  Computing and Computer-Assisted Intervention}.\hskip 1em plus 0.5em minus
  0.4em\relax Springer, 2018, pp. 347--355.

\bibitem{57}
J.~Carreira, P.~Agrawal, K.~Fragkiadaki, and J.~Malik, ``Human pose estimation
  with iterative error feedback,'' in \emph{Proceedings of the IEEE conference
  on computer vision and pattern recognition}, 2016, pp. 4733--4742.

\bibitem{58}
K.~Li, B.~Hariharan, and J.~Malik, ``Iterative instance segmentation,'' in
  \emph{Proceedings of the IEEE conference on computer vision and pattern
  recognition}, 2016, pp. 3659--3667.

\bibitem{59}
A.~Shrivastava and A.~Gupta, ``Contextual priming and feedback for faster
  r-cnn,'' in \emph{European Conference on Computer Vision}.\hskip 1em plus
  0.5em minus 0.4em\relax Springer, 2016, pp. 330--348.

\bibitem{60}
A.~R. Zamir, T.~Wu, L.~Sun, W.~B. Shen, J.~Malik, and S.~Savarese, ``Feedback
  networks,'' \emph{arXiv}, vol. abs/1612.09508, 2016.

\bibitem{48}
M.~Haris, G.~Shakhnarovich, and N.~Ukita, ``Deep back-projection networks for
  super-resolution,'' in \emph{Proceedings of the IEEE conference on computer
  vision and pattern recognition}, 2018, pp. 1664--1673.

\bibitem{43}
J.~Kim, J.~Kwon~Lee, and K.~Mu~Lee, ``Accurate image super-resolution using
  very deep convolutional networks,'' in \emph{Proceedings of the IEEE
  conference on computer vision and pattern recognition}, 2016, pp. 1646--1654.

\bibitem{44}
------, ``Deeply-recursive convolutional network for image super-resolution,''
  in \emph{Proceedings of the IEEE conference on computer vision and pattern
  recognition}, 2016, pp. 1637--1645.

\bibitem{45}
X.-J. Mao, C.~Shen, and Y.-B. Yang, ``Image restoration using convolutional
  auto-encoders with symmetric skip connections,'' \emph{arXiv preprint
  arXiv:1606.08921}, 2016.

\bibitem{46}
Y.~Tai, J.~Yang, and X.~Liu, ``Image super-resolution via deep recursive
  residual network,'' in \emph{Proceedings of the IEEE conference on computer
  vision and pattern recognition}, 2017, pp. 3147--3155.

\bibitem{47}
W.-S. Lai, J.-B. Huang, N.~Ahuja, and M.-H. Yang, ``Deep laplacian pyramid
  networks for fast and accurate super-resolution,'' in \emph{Proceedings of
  the IEEE conference on computer vision and pattern recognition}, 2017, pp.
  624--632.

\bibitem{61}
Y.~Zhang, Y.~Tian, Y.~Kong, B.~Zhong, and Y.~Fu, ``Residual dense network for
  image restoration,'' \emph{arXiv}, vol. abs/1812.10477, 2018.

\bibitem{54}
C.~Dong, Y.~Deng, C.~Change~Loy, and X.~Tang, ``Compression artifacts reduction
  by a deep convolutional network,'' in \emph{Proceedings of the IEEE
  International Conference on Computer Vision}, 2015, pp. 576--584.

\bibitem{53}
Y.~Sun, X.~Wang, and X.~Tang, ``Deep convolutional network cascade for facial
  point detection,'' in \emph{Proceedings of the IEEE conference on computer
  vision and pattern recognition}, 2013, pp. 3476--3483.

\bibitem{he2016deep}
K.~He, X.~Zhang, S.~Ren, and J.~Sun, ``Deep residual learning for image
  recognition,'' in \emph{Proceedings of the IEEE conference on computer vision
  and pattern recognition}, 2016, pp. 770--778.

\bibitem{haris2018deep}
M.~Haris, G.~Shakhnarovich, and N.~Ukita, ``Deep back-projection networks for
  super-resolution,'' in \emph{Proceedings of the IEEE conference on computer
  vision and pattern recognition}, 2018, pp. 1664--1673.

\bibitem{62}
K.~Zhang, W.~Zuo, Y.~Chen, D.~Meng, and L.~Zhang, ``Beyond a gaussian denoiser:
  Residual learning of deep cnn for image denoising,'' \emph{IEEE Transactions
  on Image Processing}, vol.~26, no.~7, pp. 3142--3155, 2017.

\end{thebibliography}
